\documentclass[aps,showpacs,prl,twocolumn]{revtex4-2}  % for review and submission 
\usepackage{siunitx}

\usepackage{amsmath}
\usepackage{mathrsfs}
\usepackage{amssymb}
\usepackage{graphicx,epstopdf,epsfig}
\usepackage{bm}
\usepackage{color}
\usepackage{times}
\usepackage{hyperref}
\usepackage{physics}
\usepackage{soul}

\newcommand{\SM}{\,\cite{SM}}

\begin{document}
%\title{Microwave Randomized Benchmarking on a neutral atom array with non-destructive readout}
\title{Randomized Benchmarking using Non-Destructive Readout in a 2D Atom Array}
\author{B. Nikolov}
\email{boyko.nikolov@strath.ac.uk}
\author{E. Diamond-Hitchcock}
\author{J. Bass}
\author{N.L.R. Spong}
\author{J.D. Pritchard}
\affiliation{Department of Physics and SUPA,University of Strathclyde, Glasgow G4 0NG, United Kingdom}

%600 characters limit
%Neutral atoms are a promising platform for scalable quantum computing, however prior demonstration of high fidelity gates or low-loss readout methods have employed restricted numbers of qubits. Using randomized benchmarking of microwave-driven single-qubit gates, we demonstrate single qubit gate errors of $8(2)\times10^{-5}$ on 225 atoms using conventional, destructive readout which exceeds the threshold for fault-tolerance. We further demonstrate suppression of measurement errors via low-loss, non-destructive and state-selective readout on 49 atoms achieving gate errors of $2(9)\times10^{-4}$ but with a 2.6$\times$ reduction in readout error which is a primary source of error in present setups.

\begin{abstract}
Neutral atoms are a promising platform for scalable quantum computing, however prior demonstration of high fidelity gates or low-loss readout methods have employed restricted numbers of qubits. Using randomized benchmarking of microwave-driven single-qubit gates, we demonstrate average gate errors of $7(2)\times10^{-5}$ on a 225 site atom array using conventional, destructive readout. We further demonstrate a factor of 1.7 suppression of the primary measurement errors via low-loss, non-destructive and state-selective readout on 49 sites whilst achieving gate errors of $2(9)\times10^{-4}$.
\end{abstract}
\maketitle

Neutral atoms have emerged as a competitive platform for scalable quantum computing \cite{saffman10,adams19,morgado21}, using arrays of optical tweezers or optical lattices to create deterministically loaded, defect-free qubit registers of atoms in up to three dimensions \cite{endres16,barredo16,barredo18,kumar18,schlosser2023} in arrays with over 1000 sites \cite{huft22}. Interactions between highly-excited Rydberg states can be exploited to perform high-fidelity two- \cite{isenhower10,maller15,zeng17,graham19,levine19,fu22,mcdonnell22} or multi-qubit gate operations \cite{levine19,pelegri21}, with the ability to realise complex connectivities using movable tweezers \cite{bluvstein22}. As well as enabling recent demonstrations of quantum algorithms \cite{graham22,bluvstein22}, this same platform can be exploited for quantum simulation \cite{ebadi21,scholl21,semeghini21} or analogue quantum computation to address practical optimization problems \cite{ebadi22,nguyen23}.

An essential requirement for future fault-tolerant scaling is the ability to perform gate operations below the error threshold \cite{knill05,steane96,devitt13}. One approach to measuring the average gate errors is randomized benchmarking, originally proposed for trapped ions \cite{Emerson2005,Dankert2009} and later adapted to a range of different hardware platforms \cite{Knill2008,Chow2009,Olmschenk10}. This method averages performance over random strings of Clifford gates to extract both gate and readout errors. Randomized benchmarking has been used to characterize microwave-driven single-qubit gates in optical lattices \cite{Olmschenk10,Lee13,Wang16} and tweezer arrays \cite{xia15,sheng18} of neutral atoms, but to date has only been demonstrated to exceed this threshold for a subset of 16 atoms \cite{sheng18}.

\begin{figure}[t!]
\includegraphics[scale=1.0]{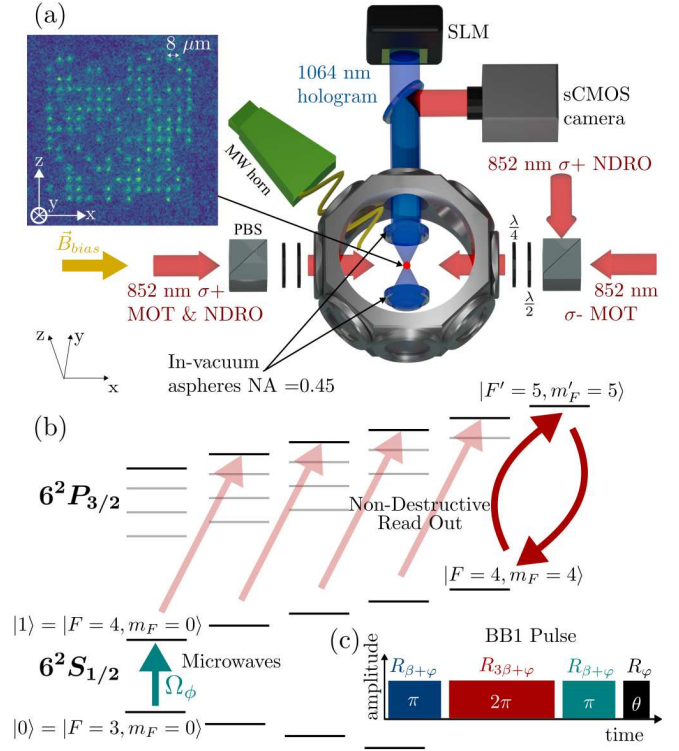}
\caption{(a) Experimental set-up for randomized benchmarking and non-destructive readout (NDRO). Atoms are trapped in a 1064~nm tweezer array and imaged onto an sCMOS camera using in-vacuum high NA lenses. NDRO is applied using a pair of counter-propagating beams aligned along the quantization axis, with microwaves emitted from a horn antenna to realise global single-qubit rotations. (b) Energy level diagram showing relevant $^{133}$Cs energy levels. Microwave gates are applied between hyperfine encoded clock qubits $\ket{0},\ket{1}$. (c) Schematic of a robust BB1 composite pulse used to implement single qubit rotation during randomized benchmarking.\label{fig1}}
\end{figure}

A second requirement for future scaling is the ability to perform scalable and high-fidelity non-destructive readout (NDRO) to suppress measurement errors arizing from the typical destructive readout scheme of ejecting atoms in qubit state $\ket{1}$ prior to imaging which cannot distinguish against atom loss during computation \cite{saffman16}. Significant progress has been demonstrated for NDRO of up to a few qubits using selective fluorescence in free-space \cite{fuhrmanek11,kwon17,MartinezDorantes2017,chow22}, a high-finesse optical cavity \cite{Bochmann10,Gehr10,Deist22}, and atomic ensembles \cite{xu21} permitting fast detection on 100~$\mu$s timescales with fidelities up to 0.999. However, quantum error correction necessitates repeated and parallel NDRO measurements of ancilla qubits in large arrays \cite{auger17,cong22}, presently only achieved using state-selective forces in a 3D lattice \cite{wu19}.

In this letter we demonstrate randomized benchmarking combined with a non-destructive readout (NDRO) across a 2D array using a low-loss technique that allows simultaneous measurement of qubit state and presence to allow post-selection against loss. First we showcase the capabilities of our experimental platform on a 225 trap site array by performing Clifford group randomized benchmarking with composite microwave (MW) pulses and a conventional, destructive readout process. Then, we implement NDRO on a 49 site array, limited by the available optical dipole trap (ODT) laser power required to create the higher trap depths needed during readout. Using this smaller array, we compare the randomized benchmarking performance obtained with conventional and non-destructive readout techniques. We demonstrate that we can efficiently transfer the atoms between the computational state $\ket{1}$, and the stretched state, where the non-destructive measurement takes place, using the NDRO beams. Finally, we discuss the limitations in the present NDRO performance and highlight experimental improvements to further reduce single atom losses. Combining this technique with long vacuum lifetime cryogenic systems \cite{Schymik2021} and atom sorting \cite{barredo16,endres16} could allow for the same atoms to be re-used for multiple experimental cycles leading to significantly faster repetition rates. This approach is also compatible with recent demonstration of mid-circuit measurement of spectator qubits in a dual-species array \cite{singh23} to achieve NDRO of ancilla qubits for quantum error correction.

%JP Updated - add missing numbers
The experiment set-up is shown schematically in Fig.~\ref{fig1}. A spatial-light-modulator (SLM) is used to generate a rectangular array of up to 15 x 15 ODTs with a 1.5~$\mu$m $1/e^2$ waist, separated by spacing of 8~$\mu$m. The ODT laser is an M Squared diode-pumped solid state laser system with a wavelength of 1064 nm, red-detuned from the $^{133}$Cs D lines, which is linearly polarized along the quantization axis. The inset of Fig.~\ref{fig1}(a) shows a typical fluorescence image obtained from stochastically-loaded single atoms in the array, imaged using a Teledyne Photometrics PrimeBSI sCMOS camera with a 47\% quantum efficiency at 852~nm and an ITO-coated, in-vacuum aspheric lens with NA=0.45 \cite{sortais07} which is also used to focus the ODT arrays. The atoms are loaded into the traps stochastically due to light-assisted collisions \cite{schlosser02}, and in each experimental run we load each site with a probability of 0.55 meaning just over half the sites are occupied. Following atom loading at a trap depth of 3 mK, a cycle of light-assisted collisions is performed to obtain single site loading and cool the atoms to 22 $\mu$K. Atoms are then prepared in the $\ket{1}=\vert F = 4,m_F = 0 \rangle$ state by optically pumping using $\pi$-polarized light driving $F=4\rightarrow F'=4$ on the D$_1$-line with a fidelity of 0.971(9), using a bias field of 6~G to define quantization. We then adiabatically lower the ODT potential to 15~$\mu$K, resulting in a final atom temperature of 8~$\mu$K. 

State-selective detection is initially performed with a single blow away beam resonant on the $^{133}$Cs D$_2$ line. The beam selectively heats atoms in $\vert F =4\rangle$ out of the trap and leaves atoms in $\vert F =3\rangle$ unperturbed. Our destructive readout errors are 1.0(1)\% for false positive detection in $\vert F=3\rangle$ and 0.5(2)\% for false negative detection in $\vert F=4\rangle$. To suppress tensor light shifts and heating from the anti-trapped excited states, during cooling, state preparation and readout the light is modulated out-of-phase with the ODT light at a frequency of 1 MHz with a 50~\% duty cycle. Further details of the experimental set-up and sequence timings are provided in the Supplementary Material \SM{}. 

\begin{figure}[t!]
\centering
\includegraphics[scale=1.0]{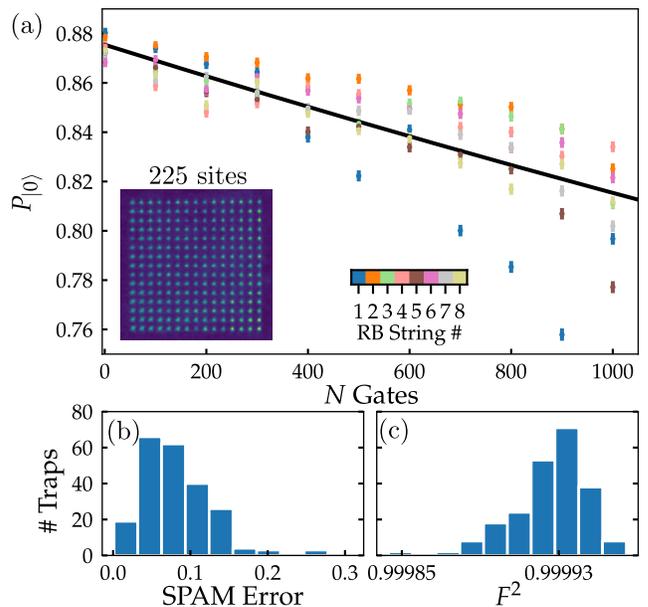}
	\caption{\label{fig2} Randomized benchmarking of 225 trap sites: (a) Array-averaged results from 8 different random gate strings with fit using equation (\ref{eq1}). Histograms showing the distribution of (b) SPAM errors and (c) average gate fidelities, $ F^2$, across the array. Errorbars represent 1 standard deviation.}
\setlength{\belowcaptionskip}{-5pt}
\end{figure}

Single qubit rotations are performed using a global microwave pulse at 9.2~GHz, derived by mixing a low phase-noise 8.95 GHz phase-locked dielectric oscillator with an agile direct digital synthesizer (DDS) operating at variable frequencies around 220 MHz. Unwanted frequency components are then removed using a bandpass filter. The filtered signal is amplified to 8~W and transmitted to the experiment using a free-space horn antenna. For the clock transition from $\ket{1}=\vert F = 4,m_F = 0 \rangle\rightarrow\ket{0}=\vert F = 3,m_F = 0 \rangle$ we obtain an averaged Rabi frequency of $\Omega/2\pi=$ 9.6~kHz. Despite the 3~cm microwave wavelength, we observe a 3~\% spatial variation in microwave Rabi frequency across the 112~$\mu$m array due to internal reflections inside the chamber. To mitigate the impact of this variation we apply microwave rotations using the BB1 composite pulse technique introduced in \cite{Wimperis1994}, and shown schematically in Fig.~\ref{fig1}(c). The BB1 protocol applies an arbitrary rotation $\theta$ around axis $\varphi$ using the four-pulse sequence $R_{\varphi+\beta}(\pi)R_{3\varphi+\beta}(2\pi)R_{\varphi+\beta}(\pi)R_{\varphi}(\theta)$ where $\beta=\cos^{-1}(-\theta/4\pi)$. This sequence reduces the sensitivity to amplitude errors to sixth order, making it robust to changes in pulse amplitude of up to 10\% \cite{wolfowicz16}. Using the microwaves we also characterize the qubit coherence time at the reduced trap depth using Ramsey interferometry, yielding an array average of $T_{2}^*$= 14.09(8)~ms \SM{}. 

To characterize the single qubit gate fidelity, we apply a randomized benchmarking approach based on uniform sampling of the 24 gates comprizing the Clifford group \cite{Dankert2009,xia15}. The specific gate set used in the experiment is obtained by building a complete set of Clifford gates from combinations of BB1 pulses implementing $R_{x,y}(\pm\pi/2)$ and $R_{x,y}(\pi)$, and is described in further detail in the Supplementary Materials \SM{}.  Rotations around $z$ axis are implemented in software as virtual gates \cite{mckay17}, where $R_z(\theta)$ corresponds to a shift in the phase of all subsequent $R_{x,y}$ pulses equivalent to rotating the coordinate axes in the Bloch sphere. This results in an average area per gate of $\langle \theta\rangle_{C_1}=2.95\pi$. 

Using this experimental method, we generate 8 different random gate strings with a maximal length reaching up to $N$=1000 gates. After applying the randomized sequence, a final correction pulse is applied to return the qubit to the $\ket{0}$ state, and the output populations are measured using 150 repeats for each data point. The total time allocated for microwave operations was fixed at 375~ms regardless of the number of gates applied in a given experimental run. The results of the randomized benchmarking on the 225-site array obtained using a destructive readout method are shown in Fig.~\ref{fig2}(a) for each of the random gate strings. Errorbars show one standard deviation. To extract the gate errors, a single fit function is applied to all datapoints using the equation
\begin{equation}
P_{\vert 0 \rangle} = \frac{1}{2} + \frac{1}{2}(1-d_\mathrm{SPAM})(1-d)^N, \label{eq1}
\end{equation}
where $N$ is the number of random gates applied, $d_\mathrm{SPAM}$ is the state preparation and measurement error, and $d$ is the average depolarization error per gate. For this conventional destructive readout approach $P_{\vert 0 \rangle}$ must be additionally scaled by the atom loss probability due to the finite trap lifetime of 9.7(8)~s, which is independently measured to be 0.93(1) \SM{}. The averaged gate fidelity is then obtained from $F^2 = 1 - d/2$ \cite{xia15}. Fig.~\ref{fig2}(b) and (c) show the distribution of SPAM error and fidelity across the array, with an array-averaged gate fidelity of $\langle F^2 \rangle = 0.99993(2)$ (corresponding to an average error of $7\times10^{-5}$) and a maximum value of {$F^2_\mathrm{max}=0.99996(3)$} at a single array site.

These results represent the highest recorded averaged single qubit gate fidelities for any quantum computing platform with over 100 qubits. A lower average microwave single qubit gate error of $4.7 \pm 1.1 \times10^{-5}$  has previously only been achieved in $4\times4$ arrays of $^{87}$ Rb atoms using magic polarization to suppress differential light shifts \cite{sheng18}. Our numbers are in good agreement with the theory from Ref.~\cite{xia15} which predicts $\langle F^2 \rangle = 1 - \langle d \rangle /2 = 1 - [1-\alpha(\langle t \rangle_{C_{1}},T_2^*)]/2$, where $\alpha = 0.5 + 0.5[1+0.95(\frac{t}{T_2^*})^2]^{-3/2}$ describes the loss of coherence due to dephasing \cite{kuhr05} and $\langle t \rangle_{C_{1}}=\langle \theta\rangle_{C_1}/\Omega$ is the average Clifford gate duration. Using the values above we predict an error of $4\times10^{-5}$, with additional sources of error likely arising from magnetic field noise or residual phase-noise in the RF electronics. 

A consequence of using BB1 pulses is the increased duration for longer gate sequences, which contributes to increased probability of loss due to background collisions and hence the high average SPAM error of 25\% due to the inability to distinguish between a real qubit error with an atom left in $\ket{1}$ at the end of the sequence and an atom being lost during the computation. 

To address this issue, we now introduce non-destructive state readout (NDRO) following the approach of \cite{kwon17, MartinezDorantes2017}. Instead of applying a resonant blow-away pulse to remove atoms in $F=4$, we perform a state-selective pulse using $\sigma^+$ polarized light on the D$_2$ $F=4\rightarrow F'=5$ transition, which optically pumps atoms in $\ket{1}$ into the $\ket{4,4}$ stretched-state to enable bright-state imaging on the cycling transition, whilst atoms in $\ket{0}$ are unaffected and remain dark. A subsequent imaging step is then performed using standard cooling and repump beams to verify the presence or absence of an atom in either hyperfine state to enable post-selection for loss.

In the experiment, NDRO is implemented using a pair of counter-propagating beams oriented along the quantization axis as shown in Fig.~\ref{fig1}(a). The beams are operated with a combined intensity equal to the saturation intensity $I_0=1.6$~mW/cm$^2$ and an average detuning of $\Delta/2\pi=-0.75\Gamma$ from the $\vert F=4,m_F=4 \rangle\rightarrow\vert F^\prime=5,m_F^\prime=5 \rangle$ transition. Where $\Gamma/2\pi=5.2$~MHz is the excited state linewidth. The beams are operated with a relative frequency difference of 600~kHz to suppress standing-wave interference. As above, light is modulated out-of-phase with the ODT to suppress AC shifts and heating. The ODT is also polarized along the quantization axis to avoid fictitious magnetic fields \cite{kwon17}. 

\begin{figure}[t!]
\centering
\includegraphics[scale=1.0]{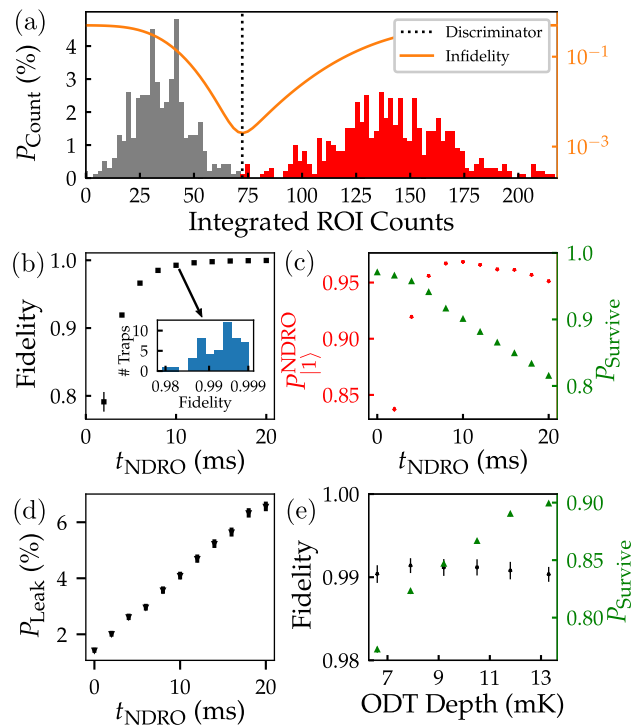}
\caption{\label{fig3} NDRO Characterization on 7x7 array at 13.3~mK trap depth. (a) Example NDRO histogram of integrated ROI counts for a single trap at $t_{\mathrm{NDRO}}=10$~ms with dashed line showing threshold that maximises detection fidelity. (b) NDRO detection fidelity vs duration, with inset showing distribution across all traps at $t_{\mathrm{NDRO}}=10$~ms. (c) NDRO detection ($P_{\ket{1}}^\mathrm{NDRO}$) and survival probabilities as a function of $t_{\mathrm{NDRO}}$. (d) Leakage probability into $\vert F=3\rangle$ during NDRO. From (b-d) we operate at $t_{\mathrm{NDRO}}=$10~ms to balance detection fidelity against errors from loss and leakage. (e) Survival and fidelity vs trap depth showing improved survival with trap depth whilst fidelity remains approximately constant.}
\setlength{\belowcaptionskip}{-5pt}
\end{figure}

Efficient NDRO requires maximizing the number of photons scattered to provide the maximal separation of bright and dark count-rates for high-fidelity state detection, whilst minimizing losses due to heating on the cycling transition or leakage into $F=3$ due to off-resonant scattering from imperfect polarization. We perform initial optimization of NDRO parameters using a 7 x 7 array, adiabatically ramping the trap to a depth of 13.3~mK for the NDRO stage (limited by the available ODT power) to suppress losses. Figure~\ref{fig3}(a) shows an example NDRO histogram obtained for 10~ms imaging duration, with red (gray) denoting counts from atoms initially prepared in the $\ket{1} (\ket{0})$ state. From these histograms we determine the threshold value that minimizes false negative or positive errors, from which we can extract the detection fidelity. In Fig.~\ref{fig3}(b-d) we present experimental data for NDRO performance as a function of readout time $t_\mathrm{NDRO}$. Fig.~\ref{fig3}(b) shows readout fidelity, (c) shows the probability of the atom surviving the readout process $P_\mathrm{Survival}$, as well as  the loss-corrected detection probability, $P_{\ket{1}}^\mathrm{NDRO}$, defined as the probability of being detected in $\ket{1}$ given the atom survived. Another important parameter is the impact of leakage into $F=3$ during the readout process. To measure the leakage rate, we perform the NDRO sequence followed by resonant blow-away to remove any atoms remaining in $F=4$ prior to the final image, with results shown in Fig.~\ref{fig3}(d). From these data we choose to operate at $t_\mathrm{NDRO}=10$~ms, offering a compromise between the requirements with a $0.900(2)$ NDRO survival probability with a state detection fidelity of $0.9926(6)$. We obtain an NDRO detection probability$P_{\ket{1}}^\mathrm{NDRO}=0.968(1)$, comparable to the measured state preparation fidelity for initializing atoms in the $\vert 1 \rangle$ state, and with a probability of leaking into $F=3$ during readout of 4.1(1)\%. The current state of the art performance for NDRO detection was reported in \cite{chow22} with $99.91\pm0.02~\%$ detection fidelity and 0.9(2) \% detection driven loss for a single Cs atom using an adaptive detection scheme using a single photon counting module.

Finally, we investigate the effect of trap depth on atom retention after NDRO which improves with increasing trap depth without compromizing discriminator fidelity as shown in Fig.~\ref{fig3}(e). The final data point corresponds to 13.3 mK trap depth which is the maximum we can achieve for this array size with our current experimental set up, with further improvements expected with more laser power.

\begin{figure}
\includegraphics[scale=1.0]{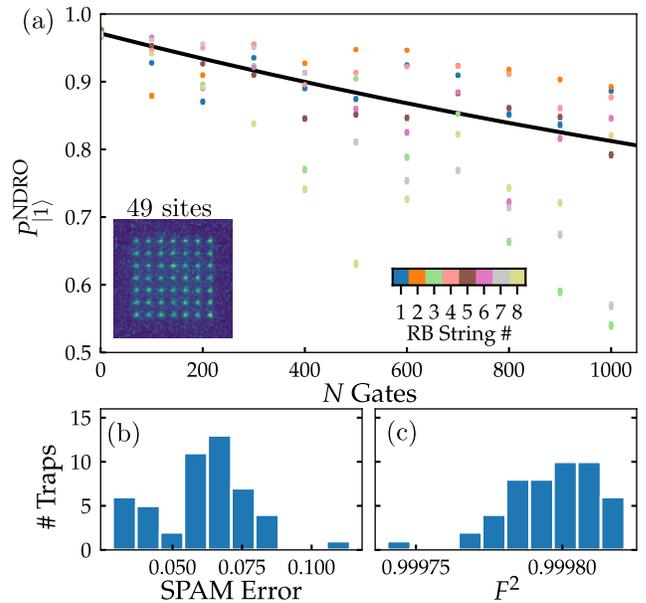}
\caption{\label{fig4}(a) Randomized benchmarking results for 49 site array readout using NDRO from $\vert1\rangle$ state, post-selected for survival. The same random  gate strings were used as for the 225 site array. Distributions of (b) average gate fidelities and (c) SPAM errors obtained with a fit using equation (\ref{eq1}) }
\end{figure}

We now apply the randomized benchmarking using the same gate sequences as above on the smaller 7x7 array with NDRO detection, performing analysis now using the loss-corrected conditional probability $P_{\ket{1}}^\mathrm{NDRO}$ with results shown in Fig.~\ref{fig4}. Using this readout scheme we achieve an array averaged fidelity of $\langle F^2\rangle=0.99978(9)$, in excellent agreement with the performance obtained in a repeated control measurement using the conventional readout sequence above with $\langle F^2\rangle=0.99978(1)$ \SM{}. The reduction in performance compared to the larger array is attributable to the finite extinction ratio of the ODT acousto-optic modulator meaning the microwave operations are performed at an increased trap depth, increasing the differential AC Stark shifts due to the ODT light from 39~Hz to 211~Hz which results in a reduction of the coherence time for the smaller array $T_2^*=12$~ms. The main improvement from the NDRO method comes from a factor of 1.7 reduction in SPAM error compared to the conventional readout with 225 sites. The SPAM error now amounts to $<5\%$ as shown in Fig.~\ref{fig4}(c). This is due to the removal of false-positives corresponding to counting lost atoms as $\ket{1}$ during the readout process, with the dominant error now the finite state preparation fidelity.

In conclusion, we have demonstrated high fidelity single qubit gate operations with an average error of $7\times10^{-5}$ by performing randomized benchmarking of global microwave pulses on an array with 225 sites using conventional, destructive detection. We have further demonstrated the ability to perform low-loss, non-destructive readout on an array of up to 49 qubits, and used NDRO method to suppress SPAM errors due to the inability to discriminate against single atom loss. Already these operations are suitable for future fault-tolerant performance, but further improvements are possible using higher microwave powers to reduce the total gate time and considering additional composite or shaped pulse sequences to suppress frequency errors \cite{wolfowicz16}.
Whilst previous experiments have achieved faster readout and lower loss for small numbers of qubits $\lesssim5$ \cite{kwon17, MartinezDorantes2017,chow22}, we have demonstrated this NDRO technique can be scaled to larger 2D arrays. 

NDRO performance is currently limited by the finite detection efficiency which limits the number of photon counts detected, and the limited trap depth during NDRO. In future, higher fidelity and reduced loss can be obtained by using an electron-multiplied CCD camera offering higher gain and quantum efficiencies $>90\%$, or using a retro-reflector to recover photons scattered away from the camera \cite{MartinezDorantes2017}, to speed up readout and suppress heating. Similarly, changing to a magic wavelength trap at 935~nm to enable trapping of both ground and excited states \cite{sheng18} both eliminates the requirement for modulating the traps whilst increasing the effective trap depth from operating closer to resonance, corresponding to approximately 5 times more traps for the same total power. 

These results extending high fidelity operations and NDRO readout to larger arrays are essential for realizing high-repetition rate digital computation by enabling atoms to be re-used after measurements, whilst also allowing for post-selection against loss. These techniques can be combined with atom sorting to allow re-loading of atoms lost during operations \cite{barredo16,endres16}, and can be adapted for applications in quantum error correction for parallel readout of ancilla qubits \cite{auger17,cong22} as required for fault-tolerant scaling.

\begin{acknowledgements}
The authors thank Andrew Daley and Arthur La Rooij for useful discussions. This work is supported by the EPSRC Prosperity Partnership \emph{SQuAre} (Grant No. EP/T005386/1) with funding from M Squared Lasers Ltd. The data presented in this work are available at \cite{nikolov23data}. 
\end{acknowledgements}

%\bibliography{../nikolov23}

\begin{thebibliography}{59}
\bibitem [{nik()}]{nikolov23data}%
  \BibitemOpen
  \bibinfo {note} {{DOI to be added in proof}}\BibitemShut {NoStop}%
\end{thebibliography}
%\bibliographystyle{apsrev4-2}


\appendix
\clearpage
\section{Supplementary Material}\label{supp}

\subsection{Experimental Sequence \& Timing Diagram}

\begin{figure}[b!]
\includegraphics[scale=1.0]{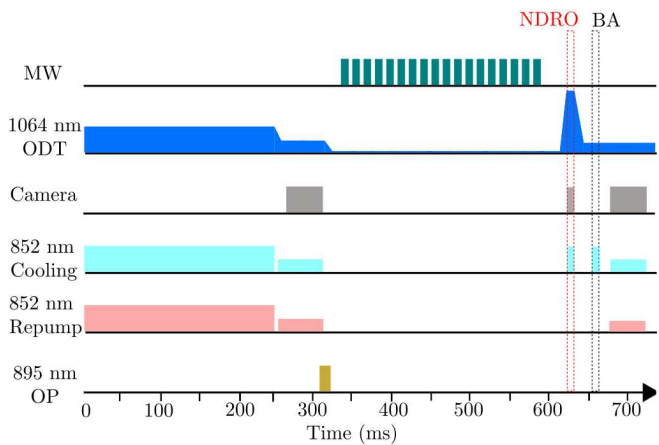}
\caption{\label{fig6}Experimental timing diagram for preparation and readout. The NDRO stage shows optional step involving ramping the trap depth up, and applying the state-selective imaging light, whilst for conventional readout imaging is preformed after a blow-away pulse (BA) to eject atoms from the trap.}
\end{figure}

Experimental timings are illustrated in Fig.~\ref{fig6}. The experiment cycle begins by loading $^{133}$Cs atoms into the arrays from a 3D magneto-optical trap (MOT) for 250~ms. After allowing the MOT to disperse, the traps are ramped to 1~mK and light-assisted collisions and polarization gradient cooling are applied to ensure that each trap site contains a single atom. An initial 40~ms image is taken using cooling and repump light to allow detection of atom loading. The atomic qubits are then optically pumped into the $\vert$F = 4,$m_F$ = 0 $\rangle$ state with a fidelity of 0.971(8) using a bias magnetic field with an amplitude of 6~G. The optical pumping fidelity is limited by the alignment of the OP light polarization vector relative to the ODT polarization and the quantization field vectors. After the optical pumping stage, the traps are adiabatically ramped down to a depth of 15~$\mu$K to further cool the atoms before applying the gate operations, achieving typical temperatures of 5-8 $\mu$K measured via release-recapture \cite{Tuchendler2008}.

Following preparation, the microwave gate pulses are triggered in sync with the 50~Hz mains to ensure repeatable magnetic field environment between different experimental runs. For the randomized gate sequences, a fixed hold time of 375~ms is used independent of gate length.

Finally, readout is performed using either conventional destructive imaging or NDRO. For conventional imaging, a resonant blow-away beam is used to eject atoms from $\vert F=4\rangle$, following by a second image to measure atom survival.  This approach yields readout errors of 1.0(1)\% for false positive detection in $\vert F=3\rangle$ and 0.5(2)\% for false negative detection in $\vert F=4\rangle$. For NDRO the trap is first ramped to a depth of around 13.3~mK, followed by applying readout light using counter-propagating beams driving the $F=4\rightarrow F'=5$ transition with $\sigma^+$ polarization to pump atoms into the stretched state cycling transition. We estimate that the efficiency of the transfer between the computational $\vert 1 \rangle$ state and the stretched state is 0.99(1) by taking a comparative measurement where we initialise the atoms directly in the state $\vert F=4, m_{F} = 4 \rangle$ using $\sigma +$ light on the D2 transition. After this optional NDRO sequence, we take a conventional image to confirm atom survival, with a 50~ms delay between exposures limited by the camera frame rate. During cooling, imaging, optical pumping and blow-away the cooling light is modulated out of phase with the ODT light at 50\% duty cycle and 1~MHz frequency to suppress heating and tensorial light shifts.

\subsection{Microwave RF Electronics}
\begin{figure}[!ht]
\includegraphics[scale=0.75]{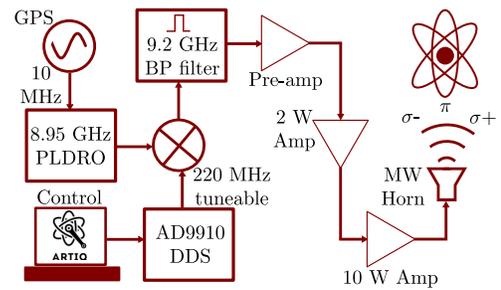}
\caption{\label{fig5}Schematic diagram of the RF chain used to apply the MW field driving single qubit rotations}
\end{figure}

The RF electronics used to produce the microwave field driving the qubit transitions are shown in Fig.~\ref{fig5}. A low phase noise Polaris phase-locked dielectric resonator oscillator (PLDRO) operating at a fixed 8.95 GHz is frequency referenced to an external 10 MHz GPS signal.  The PLDRO output is mixed with an AD9910 DDS signal and passed through a narrowband bandpass (BP) filter before being amplified in three stages to reach a total power of 8 W at 9.2 GHz. The RF power is delivered to the atoms via a microwave horn antenna directed at a vacuum chamber viewport at an arbitrary angle to the quantization axis. The distance from the horn to the tweezer array is 25~cm and the microwave radiation inside the chamber is randomly polarized with respect to the quantization axis. Different microwave-driven hyperfine state transitions are accessed by tuning the DDS frequency to be resonant with a target transition. The DDS phase and amplitude are controlled via an M-Labs artiq computer control system with the option to apply either simple rectangular or BB1 composite pulse trains as required by the experimental sequence.

\subsection{Microwave Calibration and Qubit Coherence}
\begin{figure}[t!]
\includegraphics[scale=1.0]{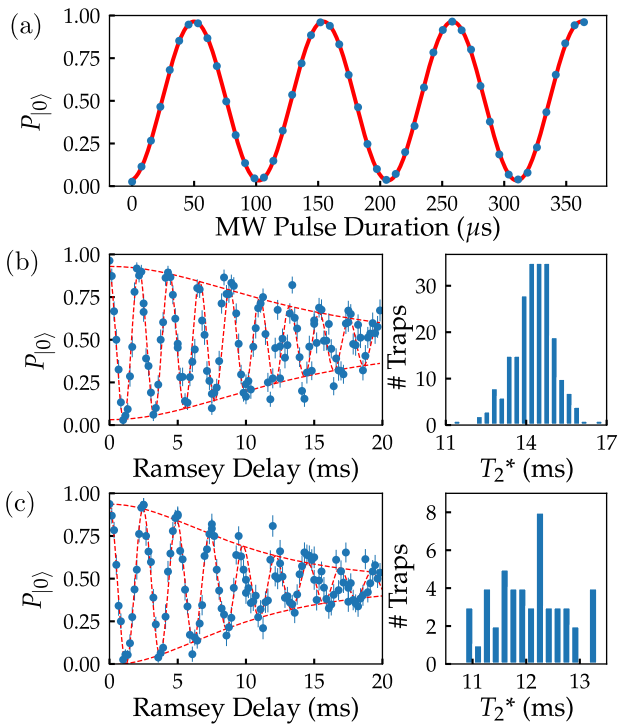}
\caption{\label{fig7}(a) MW Rabi flop on 225 site array with frequency $\Omega/2\pi$ = 9.6 kHz . (b) Single trap Ramsey fringe on 225 site array and distribution of $T_{2}^*$ times across traps. (c) Single trap Ramsey fringe on 49 site array used for NDRO and distribution of $T_{2}^*$ times across traps. }
\end{figure}

The microwave Rabi frequency is measured by applying single pulses of variable duration to the atomic array. Figure~\ref{fig7}(a) shows an array-averaged Rabi oscillation, from which we extract the average Rabi frequency $\Omega/2\pi = 9.6$~kHz, with approximately 3\% variation measured across the 225 site array. The microwave frequency is optimized using Ramsey spectroscopy, with qubit coherence times extracted by fitting the Ramsey fringe visibility as a function of delay time between two $\pi/2$-pulses with a 0.4~kHz detuning. Fitting the envelope of the oscillation to the coherence function $\alpha(t,T_2^*)$ defined above \cite{kuhr05} allows us to extract the coherence time for each site.
Figure~\ref{fig7}(b) shows an example fringe for the 225 site array, with the histogram showing the distribution of coherence times with a mean (standard deviation) of 14(0.8)~ms. Figure~\ref{fig7}(c) shows the same data for the 49 site array used for NDRO data in Fig.~4 above, with a slight reduction in the average coherence time to 12(0.6)~ms due to an increased ODT power during the microwave gate sequence.

%\begin{figure}[b!]
%\includegraphics[]{fig8.eps}
%\caption{\label{fig8}(a) Distribution of random gates sampled from Clifford Group in the eight strings shown in the main body of this letter. Gates 10 to 23 are broken down into constituent basic gates and included in this distribution. (b) Distribution of variance-weighted average gate fidelities extracted for each random gate string for the three different experimental configurations considered in this paper.}
%\end{figure}

\subsection{Implementation of Clifford Group Gate Set}
The experimental implementation of the Clifford gate set used in the experiment utilises the tuneable phase of the AD9910 DDS to implement rotations along the $x$ and $y$ axes of the Bloch sphere. In this way, an $R_y({\theta})$ pulse can be implemented by temporarily phase shifting the DDS phase by $\pi/2$ before applying a microwave pulse with area $\theta$. Since only global rotations were performed on the full 225 or 49 site arrays, the $R_z(\theta)$ gates can be implemented in software as virtual gates \cite{mckay17} by permanently offsetting the phase of all subsequent $R_x(\theta)$ and  $R_y(\theta)$ pulses by $\theta$. This can be thought of as rotating the Bloch sphere coordinate frame around the $z$ axis for all subsequent MW pulses. Since the DDS phase can be defined as leading or lagging, both positive and negative angle rotations are possible with a single MW pulse to define the first 10 Clifford gates corresponding to rotations of $\pm\pi/2$ or $\pi$ about each axis as shown in Table~\ref{table2}. The remaining 14 Clifford gates are realized with combinations of basic pulses as shown in Table \ref{table3}. The $R_i$ symbols in the final column of Table \ref{table3} are in matrix product notation, i.e applied from right to left. If single-site addressing is used, then the $R_z(\theta)$ gates cannot be implemented in the manner used in this letter. In this case, combinations of $R_x(\theta$) and $R_y(\theta$) pulses would have to be applied, similarly to the approach used in Ref.~\cite{xia15}. This is more costly in terms of pulse area. 

%Figure~\ref{fig8}(a) shows a histogram of the gates applied during each of the 8 randomised benchmark sequences, where the gate index corresponds to the primitive BB1 gate encodings defined in Table~\ref{table2}. As expected from the definition of the remaining Clifford groups in Table~\ref{table3}, $R_1$ is the most common gate as it is used in the definitions of 7/24 gates.

\begin{figure}[t!]
\includegraphics[scale=1.0]{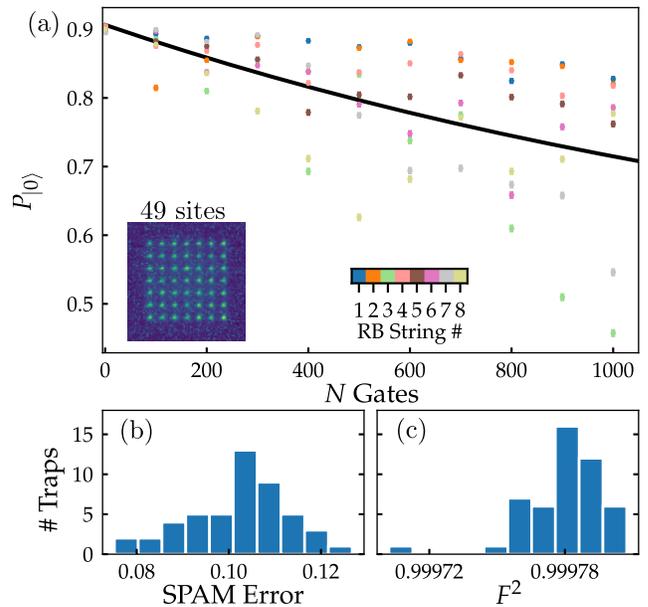}
\caption{\label{fig9}(a) Randomized benchmarking results for 49 site array readout conventional readout. (b) Distributions of  SPAM errors and (c) average gate fidelities obtained with a fit using equation (\ref{eq1}).}
\end{figure}

\subsection{Randomized Benchmarking Results}

To fit the randomized benchmarking data in Fig.~\ref{fig2}(a) of the paper using conventional readout on a 225 site array it is necessary to scale the output probability $P_{\vert 0\rangle}$ as a function of gate length from Eq.~\ref{eq1} by the total survival probability during the sequence due to the inability of this measurement technique to discriminate between atoms in $\vert 1\rangle$ and those lost due to finite trap lifetime. To determine the scaling constant we measure the baseline survival probability for the full microwave sequence in the absence of the blow-away pulse, obtaining a survival probability of 0.93(1), which is used when fitting the data to extract accurate SPAM and gate errors.

For the 49 site array, alongside the NDRO data presented in the paper we also perform randomized benchmarking on this smaller array using conventional readout to enable a direct comparison between the two techniques. These measurements are shown in Fig.~\ref{fig9}, with the fit scaled by the baseline survival measured to be 0.95(1) - this is higher than for the 225 site array due to the increase in trap depth used throughout the experiment. From the data in Fig.~\ref{fig9}(c) we obtain an averaged gate fidelity $\langle F^2\rangle=0.99978(1)$. Note no scaling factor is required for the NDRO data in Fig.~\ref{fig4} as this method already removes cases where the atoms are lost during the sequence.

\begin{table*}
% \centering
\caption{\label{table2}Basic Clifford Gate Implementation}
%\rowcolors{2}{gray!25}{white}
\begin{ruledtabular}
\begin{tabular}{cccccccc}
	Gate Index & $R_x$($\theta$) & $R_y$($\theta$) & $R_z$($\theta$) & U & Pulse Area & DDS Phase & Phase Offset\\
	%Index& & & & &Area & Current Pulse & Subsequent Pulses \\
	\hline
	0 & \it{I} & \it{I} & \it{I} & $\begin{pmatrix} 1 & 0\\ 0 & 1 \end{pmatrix}$
  & 0 & 0 & 0\\
  	1 & \it{I} & \it{I} & $\frac{\pi}{2}$ & $e^{-i\pi/4}$ $\begin{pmatrix} 1 & 0\\ 0 & \it{i} \end{pmatrix}$
  & 0 & 0 & $\frac{\pi}{2}$ \\
    	2 & \it{I} & \it{I} & $\pi$ & $-i$ $\begin{pmatrix} 1 & 0\\ 0 & -1 \end{pmatrix}$
  & 0 & 0 & $\pi$ \\
    	3 & \it{I} & \it{I} & $\text{-}\frac{\pi}{2}$ & $e^{i\pi/4}$ $\begin{pmatrix} 1 & 0\\ 0 & -\it{i} \end{pmatrix}$
  & 0 & 0 & $\text{-}\frac{\pi}{2}$\\
    	4 & \it{I} & $\frac{\pi}{2}$ & \it{I}  & $\frac{1}{\sqrt{2}}$ $\begin{pmatrix} 1 & -1\\ 1 & 1 \end{pmatrix}$
  &$\frac{\pi}{2}$ & $\text{-}\frac{\pi}{2}$ & 0 \\
    	5 & \it{I} & $\pi$ & \it{I}  & -1 $\begin{pmatrix} 0 & 1\\ -1 & 0 \end{pmatrix}$
  & $\pi$ &$\text{-}\frac{\pi}{2}$ & 0 \\
    	6 & \it{I} & $\text{-}\frac{\pi}{2}$ & \it{I}  & $\frac{1}{\sqrt{2}}$ $\begin{pmatrix} 1 & 1\\ -1 & 1 \end{pmatrix}$
  & $\frac{\pi}{2}$ & $\frac{\pi}{2}$ & 0 \\
    	7 & $\frac{\pi}{2}$  & \it{I} & \it{I}  & $\frac{1}{\sqrt{2}}$ $\begin{pmatrix} 1 & -\it{i} \\ -\it{i} & 1 \end{pmatrix}$
  & $\frac{\pi}{2}$ & 0 & 0 \\
    	8 & $\pi$  & \it{I} & \it{I}  & -\it{i} $\begin{pmatrix} 0 & 1 \\ 1 & 0 \end{pmatrix}$
  & $\pi$ & 0 & 0 \\
    	9 & $\text{-}\frac{\pi}{2}$ & \it{I} & \it{I}  & $\frac{1}{\sqrt{2}}$ $\begin{pmatrix} 1 & \it{i} \\ \it{i} & 1 \end{pmatrix}$
  & $\frac{\pi}{2}$ & $\pi$ & 0 \\
	\hline
\end{tabular}
\end{ruledtabular}
\end{table*}

\begin{table*}
\caption{\label{table3}Composite Clifford Gate Implementation}
%\rowcolors{2}{gray!25}{white}
\begin{ruledtabular}
\centering
\begin{tabular}{cccccc}
Gate Index& $R_x$($\theta$) & $R_y$($\theta$) & $R_z$($\theta$) & U &  Basic Gate Sequence\\
%& & & & & Basic Gates \\
\hline
10 & \it{I} & $\pi$ & $\frac{\pi}{2}$ & $-e^{i\pi/4}$$\begin{pmatrix} 0 & 1\\ \it{i} & 0 \end{pmatrix}$
  & $R_5$$R_1$\\
11 & $\pi$ & \it{I} & $\frac{\pi}{2}$ & $e^{i\pi/4}$$\begin{pmatrix} 0 & 1\\ -\it{i} & 0 \end{pmatrix}$
  & $R_8$$R_1$\\
12 & $\pi$ & $\frac{\pi}{2}$ & \it{I} & $\frac{-\it{i}}{\sqrt{2}}$ $\begin{pmatrix} 1 & 1\\ 1 & -1 \end{pmatrix}$
  & $R_8R_4$\\
13 & $\frac{\pi}{2}$ & \it{I} & $\frac{\pi}{2}$ & $\frac{e^{-i\pi/4}}{\sqrt{2}}$ $\begin{pmatrix} 1 & 1\\ -\it{i} & \it{i} \end{pmatrix}$
  & $R_7R_1$\\
14 & $\frac{\pi}{2}$ & $\pi$ & $\frac{\pi}{2}$ & $-\frac{e^{i\pi/4}}{\sqrt{2}}$ $\begin{pmatrix} 1 & 1\\ \it{i} & -\it{i} \end{pmatrix}$
  & $R_7R_5R_1$\\
15 & $\pi$ & $\text{-}\frac{\pi}{2}$ & \it{I} & $\frac{i}{\sqrt{2}}$ $\begin{pmatrix} 1 & -1\\ -1 & -1 \end{pmatrix}$
  & $R_8R_6$\\
16 & $\text{-}\frac{\pi}{2}$ & \it{I} & $\frac{\pi}{2}$ & $\frac{e^{-i\pi/4}}{\sqrt{2}}$ $\begin{pmatrix} 1 & -1\\ \it{i} & \it{i} \end{pmatrix}$
  & $R_9R_1$\\
17 & $\text{-}\frac{\pi}{2}$ & $\pi$ & $\frac{\pi}{2}$ & $\frac{e^{i\pi/4}}{\sqrt{2}}$ $\begin{pmatrix} 1 & -1\\ -\it{i} & -\it{i} \end{pmatrix}$
  & $R_9R_5R_1$\\
18 & $\text{-}\frac{\pi}{2}$ &$\text{-}\frac{\pi}{2}$ & \it{I} & $\frac{e^{-i\pi/4}}{\sqrt{2}}$ $\begin{pmatrix} 1 & \it{i}\\ -1 & \it{i} \end{pmatrix}$
  & $R_9R_6$\\
19 & $-\frac{\pi}{2}$ & $\frac{\pi}{2}$ & \it{I} & $\frac{e^{i\pi/4}}{\sqrt{2}}$ $\begin{pmatrix} 1 & \it{i}\\ 1 & -\it{i} \end{pmatrix}$
  & $R_9R_4$\\
20 & $-\frac{\pi}{2}$ & $\pi$ & \it{I} & $\frac{\it{i}}{\sqrt{2}}$ $\begin{pmatrix} 1 & \it{i}\\ -\it{i} & -1 \end{pmatrix}$
  & $R_9R_5$\\
21 & $\frac{\pi}{2}$ & $\text{-}\frac{\pi}{2}$ & \it{I} & $\frac{e^{i\pi/4}}{\sqrt{2}}$ $\begin{pmatrix} 1 & -\it{i}\\ -1 & -\it{i} \end{pmatrix}$
  & $R_7R_6$\\
22 & $\frac{\pi}{2}$ & $\pi$ & \it{I} & $\frac{-\it{i}}{\sqrt{2}}$ $\begin{pmatrix} 1 & -\it{i}\\ \it{i} & -1 \end{pmatrix}$
  & $R_7R_5$\\
23 & $\frac{\pi}{2}$ & $\frac{\pi}{2}$ & \it{I} & $\frac{e^{-i\pi/4}}{\sqrt{2}}$ $\begin{pmatrix} 1 & -\it{i}\\ 1& \it{i} \end{pmatrix}$
  & $R_7R_4$\\
\end{tabular}
\end{ruledtabular}
\end{table*}

\end{document}